\documentclass[journal,twoside,web]{ieeecolor}
\usepackage{jsen}
\usepackage{cite}
\usepackage{amsmath,amssymb,amsfonts}
\usepackage{algorithmic}
\usepackage{graphicx}
\usepackage{textcomp}
\usepackage{wrapfig}
\usepackage{booktabs}
\usepackage{here}
\usepackage[switch,pagewise,displaymath, mathlines]{lineno}
\usepackage{algorithm}
\renewcommand{\algorithmicrequire}{\textbf{Input:}}
\renewcommand{\algorithmicensure}{\textbf{Output:}}

\def\BibTeX{{\rm B\kern-.05em{\sc i\kern-.025em b}\kern-.08em
    T\kern-.1667em\lower.7ex\hbox{E}\kern-.125emX}}
\definecolor{abstractbg}{rgb}{0.89804,0.94510,0.83137}
\begin{document}
\title{Randomized Group-Greedy Method \\for Data-Driven Sensor Selection}
\author{Takayuki Nagata, Keigo Yamada, Kumi Nakai, Yuji Saito, Taku Nonomura
\thanks{This article was submitted on September xx. This work was supported by the Japan Science and Technology (JST) CREST Grant Number JPMJCR1763, Japan.}
\thanks{T. Nagata is with the Department of Aerospace Engineering, Tohoku University, Miyagi, JAPAN (nagata@tohoku.ac.jp). }
\thanks{K. Yamada is with the Department of Aerospace Engineering, Tohoku University, Miyagi, JAPAN (keigo.yamada.t5@dc.tohoku.ac.jp).}
\thanks{K. Nakai is with the Department of Aerospace Engineering, Tohoku University, Miyagi, JAPAN (kumi.nakai@tohoku.ac.jp).}
\thanks{Y. Saito is with the Department of Aerospace Engineering, Tohoku University, Miyagi, JAPAN (yuji.saito@tohoku.ac.jp).}
\thanks{T. Nonomura is with the Department of Aerospace Engineering, Tohoku University, Miyagi, JAPAN (nonomura@tohoku.ac.jp).}
}

\IEEEtitleabstractindextext{%
\fcolorbox{abstractbg}{abstractbg}{%
\begin{minipage}{\textwidth}%
\begin{wrapfigure}[12]{r}{3in}%
\includegraphics[width=3in]{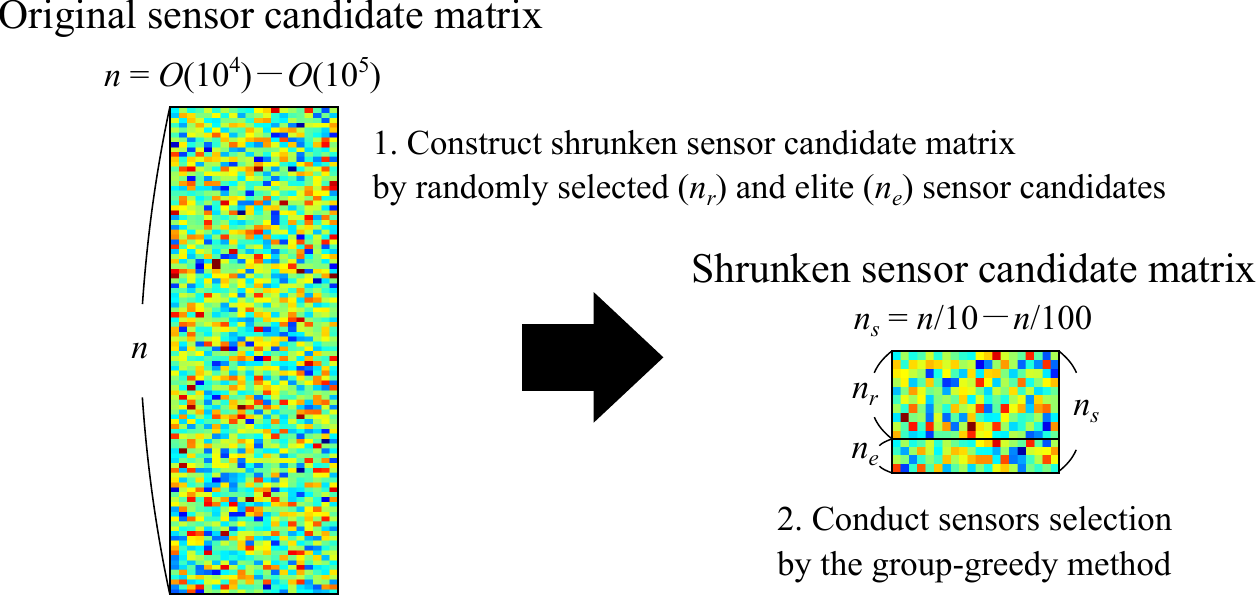}
\end{wrapfigure}%

\begin{abstract}
Randomized group-greedy methods for sensor selection problems are proposed. The randomized greedy sensor selection algorithm is straightforwardly applied to the group-greedy method, and a customized method is also considered. In the customized method, a part of the shrunken sensor candidates is selected to be the oversampled sensors by the common greedy method, and this strategy compensates for the deterioration of the solution due to shrunken sensor candidates. The proposed methods are implemented based on the D- and E-optimal design of experiments, and a numerical experiment is conducted using a randomly generated dataset. The proposed method can provide better optimization results than those obtained by the original group-greedy method when a similar computational cost is spent as for the original group-greedy method. This is because the group size for the group-greedy method can be increased as a result of the reduced sensor candidates by the randomized algorithm. 
\end{abstract}

\begin{IEEEkeywords}
Alternating direction method of multipliers, optimal design of experiment, sensor selection, sparse observation
\end{IEEEkeywords}
\end{minipage}}}

\maketitle
\section{Introduction}
\label{sec:introduction}
\IEEEPARstart{M}{easurements} of physical phenomena are an important topic in various fields. This may involve surface or volume measurements, and in most cases such measurements are performed by discretely installed point sensors. Although each sensor can only measure quantities at a particular location, full data recovery can be achieved from sparse observations by solving a linear inverse problem. It is necessary to carefully determine the position of the sensor and maximize the information obtained by sparse observations for performing the required measurement with the minimum number of sensors. 
This is referred to as the sensor placement problem (sensor selection problem), i.e., selecting the $p$ sensor locations from $n$ potential sensor locations.

Sensor placement/selection problems are formulated as a combinatorial optimization problem known as an NP-hard problem. The exact solution can be obtained by exhaustive search or global optimization techniques, such as branch and bound \cite{welch1982branch,lawler1966branch}, but these techniques can only be used for the problem of choosing a small number of sensor locations from a small number of potential sensor locations because of the high computational cost. Therefore, an approximated method that can find a suboptimal solution with a reasonable computational cost is an interesting topic.

Joshi and Boyd \cite{joshi2009sensor} applied the convex relaxation approach to sensor selection problems, and their method can obtain a global optimal solution of the relaxed problem. The computational complexity is proportional to the third power of the number of potential sensor locations. The sparsity-promoting framework based on the proximal splitting algorithm was introduced by Lin et al.~\cite{lin2013design} and Fardad et al.~\cite{fardad2011sparsity}. Their framework can obtain block-sparse feedback and observer gains as well as select actuators and sensors, and was extended by Dhingra et al.~\cite{dhingra2014admm} and Zare and Jovanovi\'{c} \cite{zare2018optimal}. Furthermore, Nagata et al.~\cite{nagata2021data} proposed a sensor selection method for a nondynamical system based on the A-optimal design of experiments and the proximal splitting algorithm. Although the computational cost was significantly reduced, it is still expensive to apply this method to a many-degree-of-freedom problem, such as data-driven sensor selection in fluid data. In addition, greedy methods will out-perform convex relaxation methods when the problem size is increased \cite{ranieri2014near,jiang2016sensor,shamaiah2010greedy}. Therefore, compared to convex relaxation methods, greedy methods are more appealing for sensor placement in a centralized context, especially for large-scale problems.

The greedy method has been studied for solving a large-scale sensor selection problem, such as the data-driven sensor selection problem \cite{manohar2018data,saito2021determinantbased,saito2020data,nakai2021effect}, and both convex relaxation and greedy methods have been extended for various purposes \cite{clark2018greedy,manohar2021optimal,astrid2008missing,barrault2004empirical,chaturantabut2010nonlinear,carlberg2013gnat,liu2016sensor,saito2020data,saito2021data,yamada2021fast,yamada2021greedy,clark2020multi,clark2020sensor}.
In a greedy algorithm, we iteratively find a new sensor location that greatly improves the objective value until the constraint is satisfied. In each step, we determine one new sensor location, and such a strategy may miss some better solutions. This is one of the difficulties of greedy methods in sensor selection problems. The performance improvement of the selected sensor subset is conducted by local optimization. For example, a 2-opt solution (i.e., no swapping of a selected and an unselected sensor has a better objective value) is searched by swapping a selected sensor and an unselected sensor \cite{joshi2009sensor}. Jiang et al.~\cite{jiang2019group} considered this issue more radically. They proposed the group-greedy (GG) method, which can obtain a better sensor subset. Their method is inspired by the beam search algorithm in the area of natural language processing \cite{tillmann2003word}. Their method iteratively reserves a group of suboptimal sensor subsets, where ''group'' indicates the top $L$ sensor subsets in the sense of a certain objective value. In this way, by considering not only the best configuration but also the suboptimal configuration, better results may be obtained than by simply pursuing only the best configuration in each step. Although the computational cost increases as the group size increases, the group-greedy method can obtain the exact solution when the group size is large enough.

In the data-driven sensor selection problem, the number of potential sensor locations becomes more than 1,000,000, as is often the case in fluid dynamics. In such a case, not only convex optimization, but also improved greedy methods, such as the group-greedy method, have a problem of computational cost, and thus, accelerated methods are required. Recently, randomized methods \cite{motwani1995randomized,halko2011finding} have been applied to convex/nonconvex problems in signal and image processing \cite{cevher2014convex,huang2014randomized,ono2019efficient}. For sensor selection problems, the convex relaxation method proposed by Joshi and Boyd \cite{joshi2009sensor} has been accelerated by Nonomura et al.~\cite{nonomura2021randomized} using the randomized subspace Newton method \cite{gower2019rsn}. Hashemi et al.~\cite{hashemi2020randomized} proposed a randomized greedy method that selects sensors for state estimation in large-scale linear time-varying dynamical systems. They provided a performance guarantee for the proposed algorithm and demonstrated that the randomized method is superior to the common greedy and semidefinite problem relaxation methods in terms of computational time while selecting the same or better sensor subset.

In the present study, we propose a randomized group-greedy method and customized group-greedy method. The sensor selection problem is such that $p$ sensors are selected from among $n$ potential sensors. Each sensor gives an observation vector $\mathbf{y}$ of a linear function of latent state variables $\mathbf{z}$ superimposed with independent identically distributed zero-mean Gaussian random noise. The main contributions of the present paper are summarized as follows:
\begin{itemize}
    \item The computational cost of the group-greedy method is significantly reduced by introducing the randomization technique.
    \item The proposed method can solve a large-scale problem with almost the same computational time as the common greedy method, and a better sensor subset can be obtained in terms of the objective value.
    \item Because of reduced computational cost, a search with a larger group size (deeper exploration), as compared to the original group-greedy method, can be conducted.
    \item By adding elite sensor candidates obtained by oversampling with a low-cost method, such as the common greedy method, to the random sketch of the sensor candidates, the performance of the selected sensor subset is further improved in exchange for a slight increase in computational cost.
\end{itemize}
\section{Problems and Algorithms}\label{sec:probandalg}
\subsection{Sensor Selection Problems}
We consider the sparse measurement of state variables including uniform independent Gaussian measurement noise $\mathbf{v}=\mathcal{N}(\mathbf{0},\sigma^2\mathbf{I}) \in\mathbb{R}^p$ generated by latent variables $\mathbf{z}\in\mathbb{R}^{r}$. The following equations are considered:
\begin{align}
    \mathbf{y}&=\mathbf{HUz}+\mathbf{v}, \nonumber \\
              &=\mathbf{Cz}+\mathbf{v},
    \label{eq:observation}
\end{align}
where $\mathbf{y}\in\mathbb{R}^p$ is the observation vector, $\mathbf{H}\in\mathbb{R}^{p\times n}$ is the sensor location matrix, and  $\mathbf{U}\in\mathbb{R}^{n\times r}$ is the sensor candidate matrix. In addition, $\mathbf{C}\in\mathbb{R}^{p\times r}$ is the measurement matrix, which is the product of the sensor location matrix and the sensor candidate matrix $\mathbf{C}=\mathbf{HU}$. Here, the variables $p$, $n$, and $r$ indicate number of sensors, the number of potential sensor locations, and the number of latent variables, respectively. The sensor location matrix $\mathbf{H}$ is a sparse matrix that indicates sensor locations. Each row vector of $\mathbf{H}$ is a unit vector, and the locations of unity in each row vector correspond to the activated sensor locations chosen from $n$ potential sensor locations. The locations of the activated sensors are selected based on the optimal design of experiments. The E- and D-optimality criteria that are frequently used are considered to be as follows:
\begin{align}
    &f_{\mathrm{E}}\,=\,\left\{\begin{array}{cc}
        \lambda_{\mathrm{min}}\,\left(\mathbf{CC}^{\mathsf T}\right) & p\le r \\
        \lambda_{\mathrm{min}}\,\left(\mathbf{C}^{\mathsf T}\mathbf{C}\right) & p>r
    \end{array}\right.\label{eq:obj_eig}, \\
    &f_{\mathrm{D}}\,=\,\left\{\begin{array}{cc}
        \mathrm{det}\,\left(\mathbf{CC}^{\mathsf T}\right) & p\le r \\
        \mathrm{det}\,\left(\mathbf{C}^{\mathsf T}\mathbf{C}\right) & p>r
    \end{array}\right.\label{eq:obj_det}.
\end{align}
Maximization of $f_\mathrm{D}$ and $f_\mathrm{E}$ corresponds to minimization of the volume of the confidence ellipsoid and the worst-case error variance, respectively. Sensor selection by the greedy method based on these objective functions is described in \cite{saito2021determinantbased,nakai2021effect}. The function $f_{\rm E}$ is not a submodular function \cite{nakai2021effect}, and thus optimization using the common greedy method does not work well. Therefore, the benefits of introducing the group strategy are great. On the other hand, the objective function $f_{\rm D}$ is a monotone submodular function \cite{saito2021determinantbased} and can be optimized with a good approximation guarantee even by using the common greedy method. Therefore, the advantages of the GG method in the case of D-optimality-based methods are fewer than that in the case of E-optimality-based methods.

\subsection{Group-Greedy Method}
The group-greedy method was proposed by Jiang et al. \cite{jiang2019group}. This method can be applied to all methods based on the greedy algorithm, and the performance of these method in terms of the objective value improves. In the common greedy method, the best location is repeatedly selected in each step 
until a constraint, such as the number of selected sensors $k$, is reached a predefined number $p$. In this case, the obtained sensor subset might be a suboptimal configuration because the obtained sensor subset is the set of the optimal solutions for each divided problem. The group-greedy method considers the $k+1$th sensor not only in the current optimal configuration, but also in suboptimal configurations. Although the computational cost increases, the result should be improved by increasing the group size $L$. Here, the group size $L$ is the number of stored suboptimal configurations.

\subsection{Randomized Group-Greedy Method}
The group-greedy method can obtain better optimization results, but the computational cost becomes a critical issue for large-scale problems, which have more than $\mathcal{O}(10^4)$ potential sensor locations. In the present study, a randomization technique is introduced to significantly reduce the computational cost. In addition, the proposed method can conduct optimizations with a larger group size than the original group-greedy method 
because of the reduced computational cost.

The sparse sampling assumes a low rankness of the sampling target. On the other hand, the number of potential sensor locations is large. In particular, the number of potential sensor locations might reach more than $\mathcal{O}(10^5)$ in the data-driven sensor selection. In such a case, there is a large number of potential sensor locations, but only a small number of sensors is required 
for full data reconstruction. Therefore, reduction of the sensor candidate matrix by projecting onto randomized subspace is effective.

In the proposed method, shown in Alg.~\ref{alg:RGG}, the shrunken sensor candidate matrix is generated before selecting sensors by the group-greedy method, as shown in Alg.~\ref{alg: group greedy}. The randomized group-greedy (RGG) method conducts sensor selection by the group-greedy method in the shrunken sensor candidates, which is the random subset $\mathcal{S}_s$ of the original sensor candidates $\mathcal{S}$. The number of sensor candidates is reduced from $|\mathcal{S}|=n$ to $|\mathcal{S}_{\rm s}|=n_{\rm s}$ (the reduction ratio $n_{\rm s}/n$ is 1/10 or 1/100), and the number of evaluations is reduced from $Ln$ to $Ln_{\rm s}$, where $n_{\rm s}$ is the size of the shrunken sensor candidate set generated by random sampling. 
However, valuable locations are possibly truncated when generating a random subset of the sensor candidate matrix, and the optimization result might be degraded. 
Therefore, an elite strategy that adds the location selected by low-cost methods to a random subset of the sensor candidate matrix is introduced (elite-and-randomized-group-greedy (ERGG) method). In particular, the sensor set selected by the group-greedy method often contains the sensor locations selected by the common greedy method. An improved strategy is thus considered here by including a subset of elite sensors as sensor candidates, and the performance degradation due to the shrunken sensor candidate matrix is minimized. In this method, the shrunken sensor candidate $\mathcal{S}_{\rm s}$ becomes $\mathcal{S}_{\rm r}\sqcup\mathcal{S}_{\rm e}$, where $\mathcal{S}_{\rm e}$ is the elite sensor candidate, and the size of each set is $|\mathcal{S}_{\rm s}|=n_{\rm s}$, $|\mathcal{S}_{\rm r}|=n_{\rm r}$, and $|\mathcal{S}_{\rm e}|=n_{\rm e}$. Note that $n_{\rm r}$ in the ERGG method is set so that $n_{\rm s}$ becomes the same as that for the RGG method in the present numerical experiments. Although elite sensor candidates can be selected in various ways, the candidates are selected using the common greedy method in order to simplify the discussion in the present study. 

\begin{algorithm}
\caption{Elite and randomized group-greedy method} \label{alg:RGG}
\begin{algorithmic}
\renewcommand{\algorithmicrequire}{\textbf{Input:}}
\renewcommand{\algorithmicensure}{\textbf{Output:}}
\REQUIRE $\mathbf{U}\in\mathbb{R}^{n \times r},\, p\in\mathbb{N},\, L\in\mathbb{N},\, \mathcal{S}_{\rm e}$
\ENSURE  Indices of the top subset for $p$ sensor positions $\mathcal{S}_{p, 1}$
    \STATE Set $k \leftarrow 1$, $\mathcal{S}_{0}\leftarrow \emptyset$
	\FOR{ $k\leq p$ }
	\IF{ $k=1$ }
	\STATE Set candidate $\mathcal{S}_{\rm s} \leftarrow \mathcal{S} := \{ 1,\, \hdots,\, n \}$
	\STATE Alg.~\ref{alg: group greedy} with $\mathbf{U}$, $\mathcal{S}_{0}$, $\mathcal{S}_{\rm s}$ and $L$\\
	\ELSE

	\FOR{ $l \in \{1,\, 2,\, \hdots L\}$ }
    \STATE Set $\mathcal{S_{\rm r}}$ by preliminary random selection from $\mathcal{S}_{n}\backslash\mathcal{S}_{e}$
    \STATE Set combined candidate $\mathcal{S}_{\rm s} \leftarrow \mathcal{S}_{\rm r} \sqcup \mathcal{S}_{\rm e}$
    \STATE Alg.~\ref{alg: group greedy} with $\mathbf{U}$, $\mathcal{S}^{(k-1,l)}$, $\mathcal{S}_{\rm s}$ and $L$
	\ENDFOR
    \STATE Store all $\mathcal{F} $ and $\mathcal{T}_{k} $
	\ENDIF
    \STATE Compare $\mathcal{F} $ and store $L$-best sensor subsets $\mathcal{S}^{(k,l)} \, (l \in \{ 1, \hdots, L \})$ from $\mathcal{T}_{k}$ eliminating duplication
	\STATE Set $k \leftarrow k+1$
	\ENDFOR 
	
    Select best sensor subset $\mathcal{S}_{p,1}$ from $\mathcal{S}^{(k,l)}$
\end{algorithmic}
\end{algorithm}

\begin{algorithm}
\caption{$L$-best greedy search for the $k$-th sensor} \label{alg: group greedy}
\begin{algorithmic}
\renewcommand{\algorithmicrequire}{\textbf{Input:}}
\renewcommand{\algorithmicensure}{\textbf{Output:}}
\REQUIRE $\mathbf{U}\in\mathbb{R}^{n \times r}, \, \mathcal{S}^{(k-1)} = \{ i_{1},\, i_{2},\, \hdots,\, i_{(k-1)} \},\,\mathcal{S}_{\rm s} ,\, L \in \mathbb{N}$ 
\ENSURE  $\mathcal{F}\in\mathbb{R}^{L}$, $\mathcal{T}_{k}\in\mathbb{R}^{L \times k}$
    \STATE Calculate objective values $f\left(\mathcal{S}^{(k-1)}\cup i \right)$ for $\forall\, {i\, \in\, \mathcal{S}_{\rm s}\, \backslash\, \mathcal{S}^{(k-1)}}$\\
    \STATE    $\mathcal{F} \leftarrow$ Best $L$ objective values of $f$\\ 
    \STATE    $\mathcal{T}_{k} \leftarrow $ Corresponding $L$ subsets of sensor location 
\end{algorithmic}
\end{algorithm}    

The computational complexities of the proposed and previously proposed methods are shown in Table~\ref{tab:complexity}.

\begin{table}
\caption{Comparison of computational complexity (D-optimality-based methods).}
\label{table}
\setlength{\tabcolsep}{3pt}
\begin{tabular}{|p{120pt}|p{80pt}|}
\hline
Method& 
Complexity \\
\hline
Common greedy& 
$\mathcal{O}\left(pnr^2\right)$\\
Group-greedy& 
$\mathcal{O}\left(Lpnr^2\right)$\\
Randomized group-greedy& 
$\mathcal{O}\left(Lpn_{\rm s}r^2\right)$\\
Elite and randomized group-greedy& 
$\mathcal{O}\left(Lpn_{\rm s}r^2+n_{\rm e}nr^2\right)$\\
\hline
\end{tabular}
\label{tab:complexity}
\end{table}

\section{Results and Discussion}\label{sec:results}
The performance of the proposed methods was evaluated by applying those methods to randomly generated sensor candidate matrices $\mathbf{U}\in\mathbb{R}^{n\times r}$ in which the entries follow a normal distribution of $\mathcal{N}(0,1)$. The number of sensor candidates was set to $n=10,000$, and the number of latent variables was set to $r=10$. The computation for each condition was conducted 500 times with a different sensor candidate matrix, and average values of the objective functions (E- and D-optimality criteria) and computational time were evaluated. The performance of the proposed methods was compared with that for common greedy \cite{nakai2021effect,saito2021determinantbased} and original group-greedy (GG) methods \cite{jiang2019group}. 

\subsection{Comparison with the Previously Proposed Method}
Numerical experiments with two parameter settings were conducted. For the first case, the parameters for randomized group-greedy methods (RGG and ERGG) were set so that the computational cost would be the same as that for the common greedy method ($L_{\rm RGG}n_{\rm s}=L_{\rm ERGG}n_{\rm s}=n$). In this case, the size of the shrunken sensor candidate matrix was set to $n_{\rm s}=1000$, and the group size for the GG, RGG, and ERGG methods was set to $L=10$. For the second case, the parameters for the randomized group-greedy methods were set so that the computational cost would be the same as that for the GG method at $L_{\rm GG}=10$ ($L_{\rm RGG}n_{\rm s}=L_{\rm ERGG}n_{\rm s}=L_{\rm GG}n$). In this case, the size of the shrunken sensor candidate matrix and the group size for the RGG and ERGG methods were set to $n_{\rm s}=1000$ and $L=100$, respectively. The number of elite sensor candidates for the ERGG method was fixed at $n_{\rm e}=100$ (i.e., $n_{\rm r}=900$). The influence of the parameter on the performance of the proposed methods will be discussed in Section~\ref{sec:param}.

\begin{figure}[!tbp]
\centering
\includegraphics[width=3in]{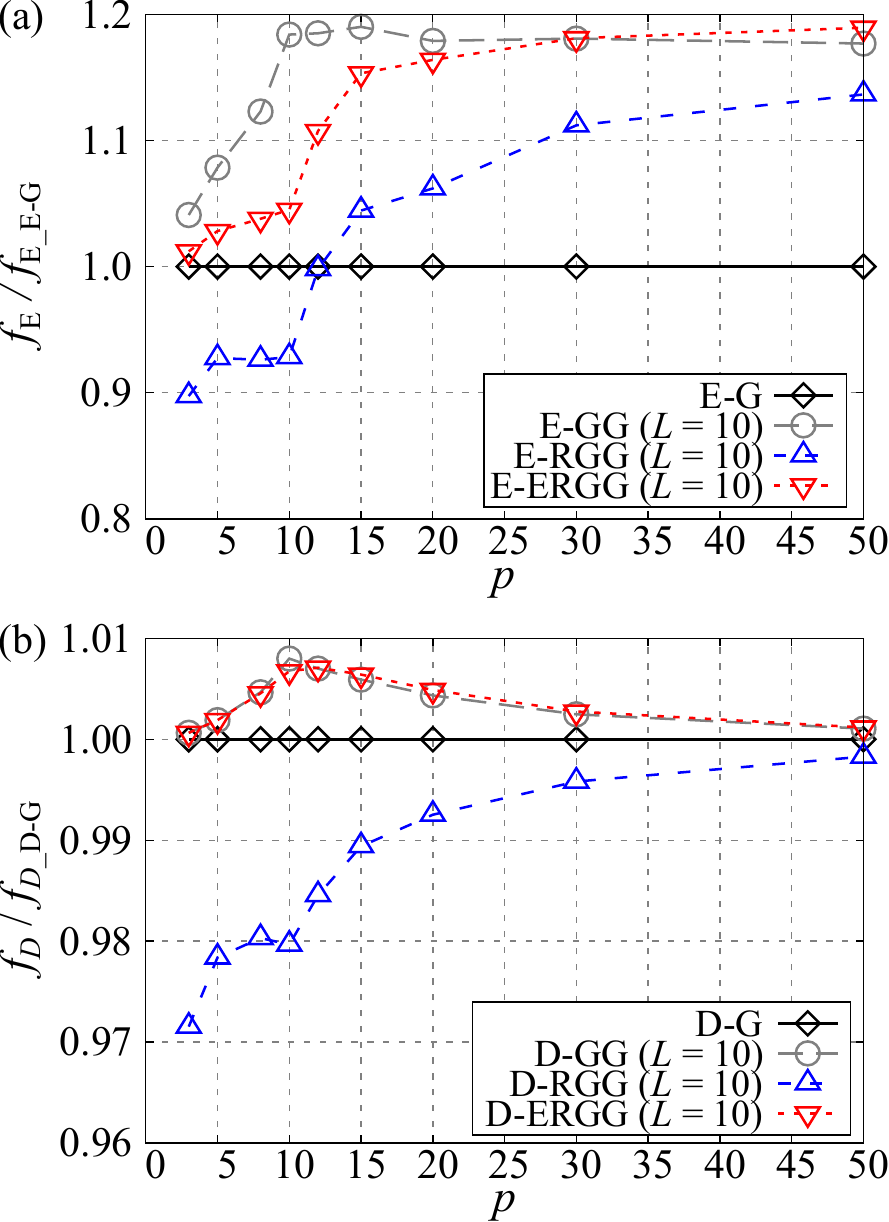}
\caption{Objective values with respect to the number of sensors when the computational cost of the randomized methods is the same as the common greedy method ($L_{\rm RGG}n_{\rm s}=n$). (a) E-optimality-based methods; (b) D-optimality-based methods.}
\label{fig:obj-G} 
\end{figure}

Objective values are compared in Fig.~\ref{fig:obj-G}. The performance of the GG method is the best in most conditions of $p\leq50$. Note that the number of evaluations for the GG method is $Ln=100,000$, but it is 10 times smaller for the RGG and ERGG methods.
The objective values obtained by the RGG methods are decreased compared to those obtained by the common greedy methods due to the shrunken sensor candidate matrix. The performance of the RGG methods is improved by introducing the elite sensor candidates (ERGG method), and the objective values obtained by the randomized methods are asymptotic to that obtained by the GG method for larger $p$.

There are several differences in the characteristics of the obtained objective values between the E-optimality-based and D-optimality-based methods. For the RGG methods, the performance of the E-RGG method is superior to that of the E-G method in oversampling conditions, but the objective value obtained by the D-RGG method never exceeds the value obtained by the D-G method at $p\leq 50$. This is because $f_{\rm D}$ is a monotone submodular function \cite{saito2021determinantbased} and can be optimized with a good approximation guarantee, even that obtained using the common greedy method. The objective value obtained by the D-RGG method does not exceed that obtained by the D-G method for $p\leq 50$, and thus the D-RGG method is not effective. However, degradation of the solution due to the shrunken sensor candidate matrix is compensated for in the D-ERGG method, and objective values obtained by the D-ERGG method become approximately the same as those obtained by the D-GG method. On the other hand, the function $f_{\rm E}$ is not a submodular function \cite{nakai2021effect}, and thus optimization using the common greedy method does not work well. Therefore, the benefits of introducing the group strategy are great \cite{jiang2019group}, and the performance of the E-RGG method in oversampling conditions is better than that of the E-G method, even though the sensor candidate matrix is shrunken. The difference in the relative performance between the E-G and E-RGG methods in undersampling and oversampling conditions is caused by the nature of the E-G method. The performance of the E-G method is relatively high in the undersampling conditions, but it becomes rapidly worse in oversampling conditions \cite{nakai2021effect}. Hence, the performance of the E-RGG method is inferior in undersampling conditions due to the shrunken sensor candidate matrix and is superior in oversampling conditions because of the benefit brought about by the group strategy, even though the sensor candidate matrix is shrunken. The performance of the E-ERGG method is even higher, and the objective values obtained by the randomized methods are asymptotic to those obtained by the E-GG method at $p>20$. However, unlike the D-optimality-based method, the E-ERGG method is significantly inferior to the E-GG method in undersampling conditions. This is because the performance of the E-G method used for selecting elite sensor candidates is poor due to the lack of submodularity of objective function $f_{\rm E}$.

The difference in the trend between the E-optimality-based method and the D-optimality-based method is caused by the presence or absence of submodularity in the objective functions. This tendency is the same in the discussions that follow.

\begin{figure}[!tbp]
\centering
\includegraphics[width=3in]{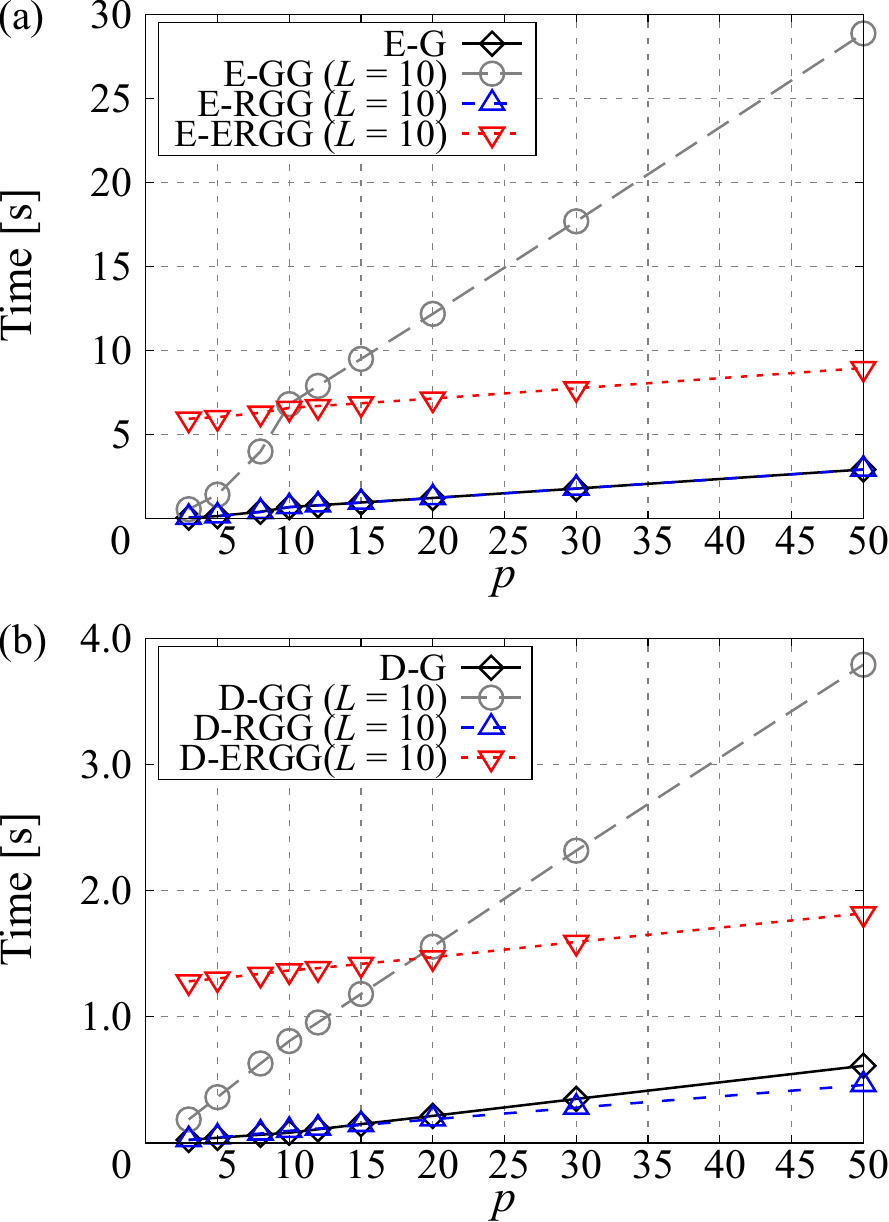}
\caption{Computational time with respect to the number of sensors when the computational cost for the RGG method is the same as that for the common greedy method ($n=L_{\rm RGG}n_{\rm s}$). (a) E-optimality-based methods; (b) D-optimality-based methods.}
\label{fig:time-G} 
\end{figure}

Comparison of the computational time is shown in Fig.~\ref{fig:time-G}. The computational time for the GG method rapidly increases as the number of sensors to be selected increases. On the other hand, the computational time for the RGG method is the same as that for the common greedy method because the number of evaluations is the same. Although the computational time is the same as that for the common greedy method, the RGG method can obtain a better solution in oversampling conditions when the objective function is the value of the E-optimality. The increase in the computational time for the ERGG method with respect to the number of sensors to be selected is on the same level as that for the common greedy method, but the cost for selecting elite sensor candidates is added. By allowing this additional cost, a significant performance improvement can be obtained, and the computational time is shorter than that for the GG method when the number of sensors is large.

\begin{figure}[!tbp]
\centering
\includegraphics[width=3in]{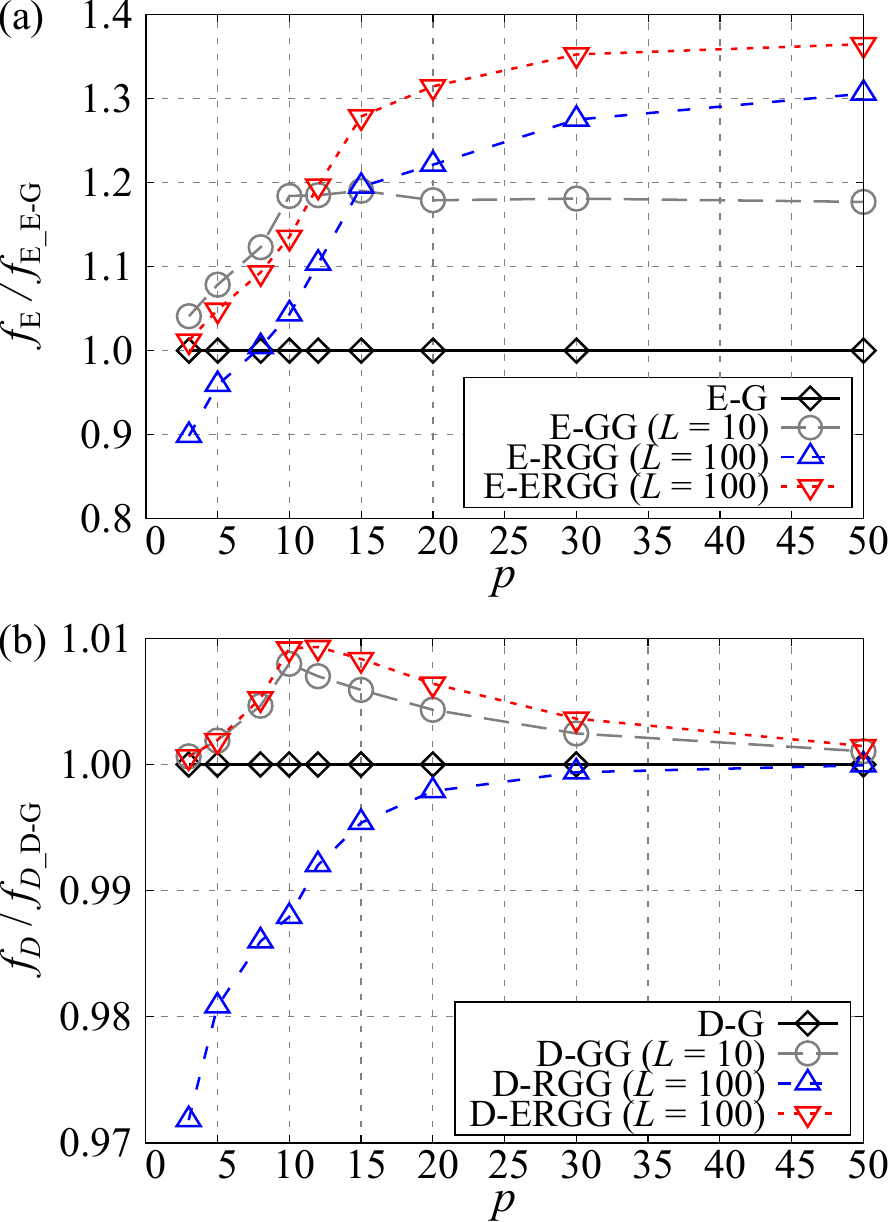}
\caption{Objective values with respect to the number of sensors when the computational cost for the RGG method is the same as that for the original group-greedy method ($L_{\rm GG}n=L_{\rm RGG}n_{\rm s}$). (a) E-optimality-based methods; (b) D-optimality-based methods.}
\label{fig:obj-GG} 
\end{figure}

\begin{figure}[!tbp]
\centering
\includegraphics[width=3in]{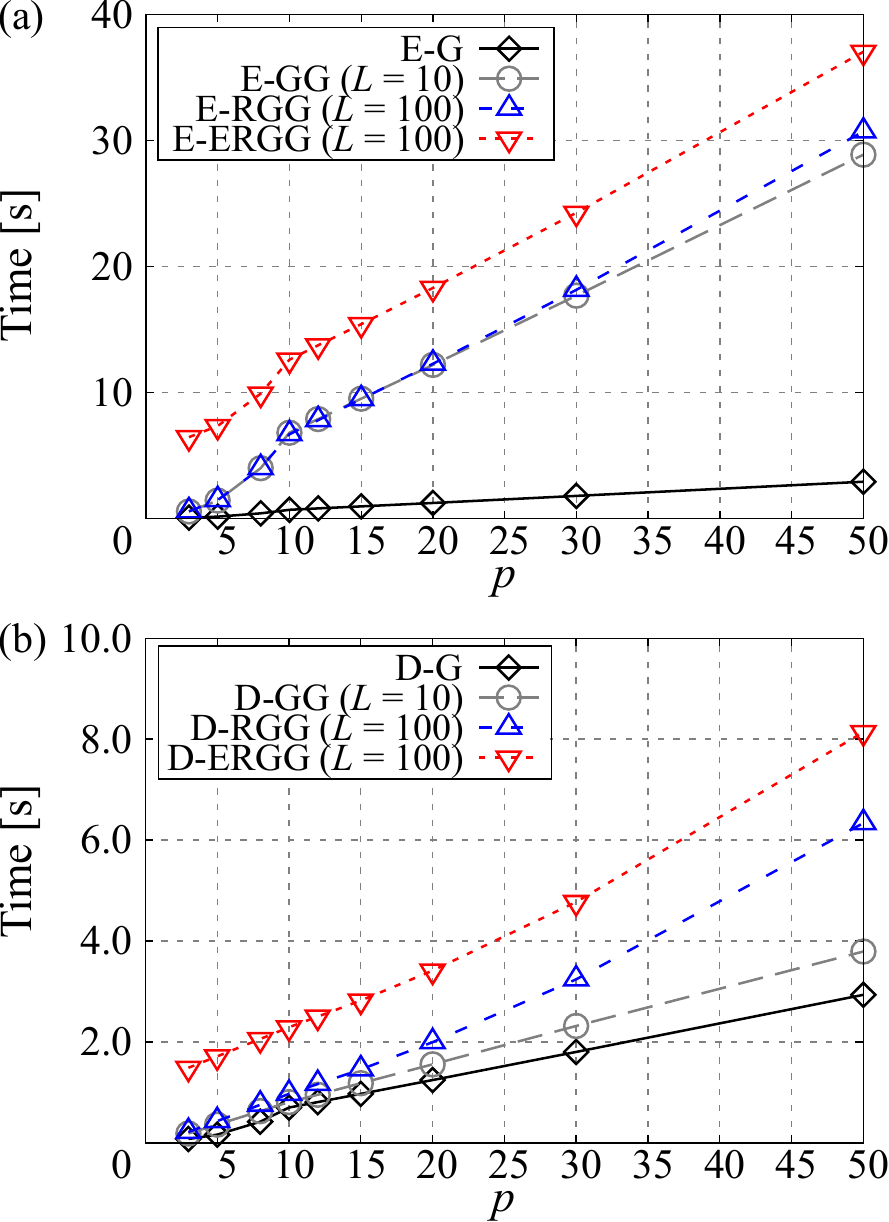}
\caption{Computational time with respect to the number of sensors when the computational cost for the randomized method is the same as that for the original group-greedy method ($L_{\rm GG}n=L_{\rm RGG}n_{\rm s}$). (a) E-optimality-based methods; (b) D-optimality-based methods.}
\label{fig:time-GG} 
\end{figure}

Comparisons of the objective value and the computational time when the number of evaluations in the RGG and ERGG methods is the same as that in the GG method are presented in Figs.~\ref{fig:obj-GG} and \ref{fig:time-GG}, respectively. The numbers of evaluations for the GG, RGG, and ERGG methods are $L_{\rm GG}n=L_{\rm RGG}n_{\rm s}=L_{\rm ERGG}n_{\rm s}$. The objective value obtained by the E-RGG method is superior to that obtained by the E-GG method at $p\geq15$, even though the computational cost is the same. The objective value obtained by the E-ERGG method is further improved, in exchange for a slight increase in the computational time. At $p\leq15$, on the other hand, the performance of the E-RGG method is degraded compared to the E-GG method, even though the number of evaluations is the same. This is because the valuable location is missed by shrinking the sensor candidates in the case of the E-RGG method. Degradation of the solution due to the shrunken sensor candidate matrix can be reduced by using the E-ERGG method. However, the common greedy method is used for selecting the elite sensor candidates in the present study on the E-ERGG method, and thus the objective value is still degraded compared with that for the E-GG method.

The objective value obtained by the D-RGG method is smaller than not only that obtained by the D-GG method but also that obtained by the D-G method. despite the same number of evaluations as the D-GG method. Although the difference in the objective values obtained by the D-GG and D-ERGG methods is smaller than that obtained by the E-GG and E-ERGG methods, the objective value obtained by the D-ERGG method is larger than that for the D-GG method, particularly in oversampling cases. The computational time for the D-RGG method is clearly larger than that for the D-GG method due to the larger $L$. This is because the computational cost for the D-optimality-based greedy methods is low, and the cost specific to the group-greedy method, such as sorting, has a larger impact.

\subsection{Influence of Parameters on Performance} \label{sec:param}
\begin{figure}[!tbp]
\centering
\includegraphics[width=3in]{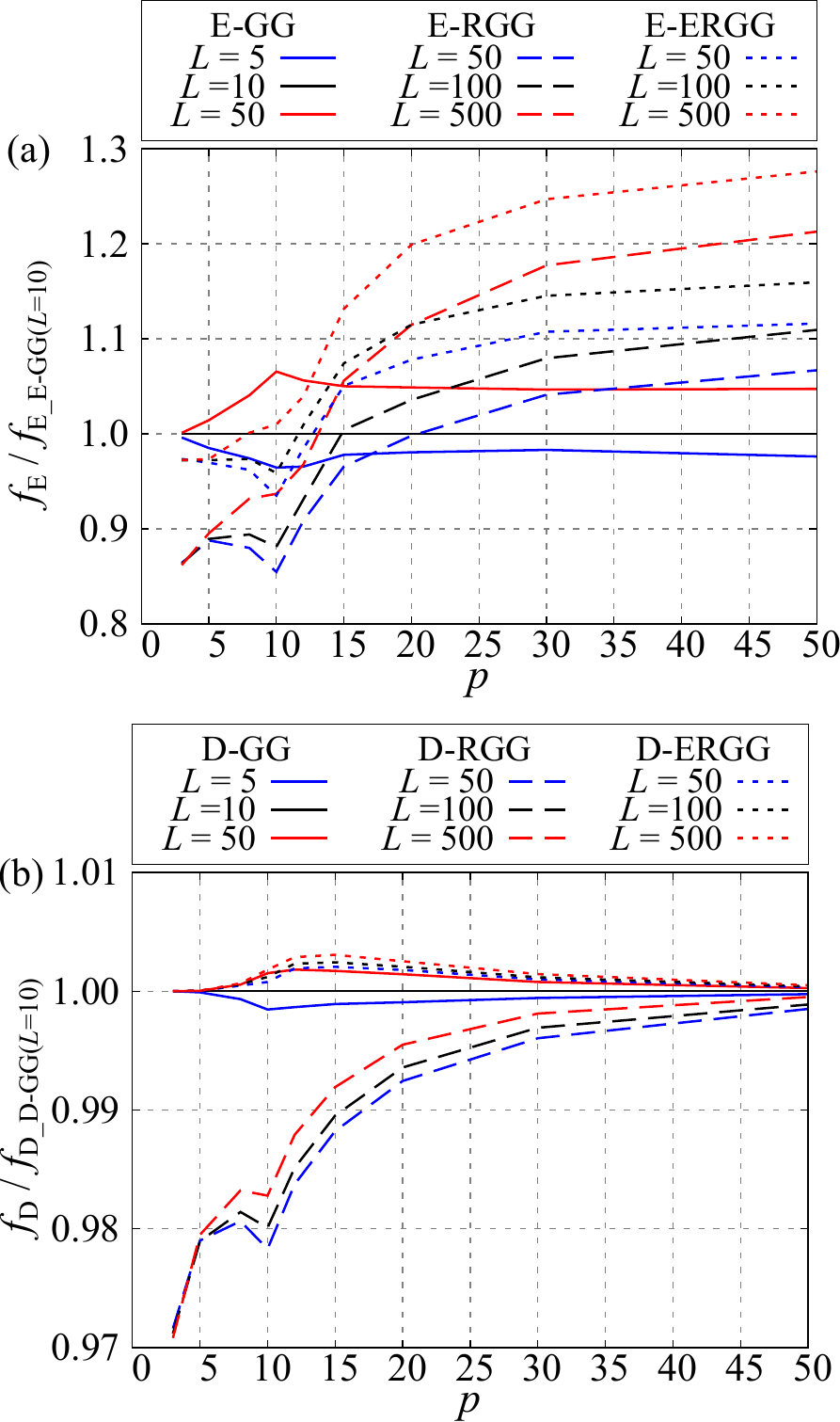}
\caption{Effects of the group size $L$ on objective values. The shrink ratios for the RGG and ERGG methods are $n_s/n=0.1$, and the number of evaluations of the RGG and ERGG methods are the same as that of the GG method. (a) E-optimality-based methods; (b) D-optimality-based methods.}
\label{fig:eig-L} 
\end{figure}

The proposed method has three parameters, which are the group size $L$, the size of the shrunken sensor candidate matrix $n_{\rm s}$, and the number of elite sensor candidates $n_{\rm e}$. The effect of the group size on the objective value is shown in Fig.~\ref{fig:eig-L}. The number of evaluations for the GG, RGG, and ERGG methods are the same ($L_{\rm GG}n=L_{\rm RGG}n_{\rm s}=L_{\rm ERGG}n_{\rm s}$) for the same line colors in Fig.~\ref{fig:eig-L}. The shrink ratio for the RGG and ERGG methods is $n_{\rm s}/n=0.1$, where $n=10,000$.

In the case of E-optimality-based methods, there is no large effect of $L$ on the objective value in each method at $p=3$, but the objective values are different between the E-GG, E-RGG, and E-ERGG methods. The influence of $L$ on the objective value appears at larger $p$ in undersampling conditions. Although the objective values are different for each method, the increase of the objective value by increasing $p$ is almost the same when the number of evaluations is the same ($L_{\rm GG}n=L_{\rm RGG}n_{\rm s}=L_{\rm ERGG}n_{\rm s}$). As discussed in the previous sections, the performance of the E-RGG method is degraded compared to the E-GG method due to the shrunken sensor candidate matrix. In particular, although the number of evaluations for the E-RGG method is ten times larger than that for the E-GG method, the objective value obtained by the E-RGG method with $L=500$ is smaller than that obtained by the E-GG method with $L=5$. Therefore, the number of sensor candidates is more important than the number of evaluations and the group size.

\begin{figure}[!tbp]
\centering
\includegraphics[width=3in]{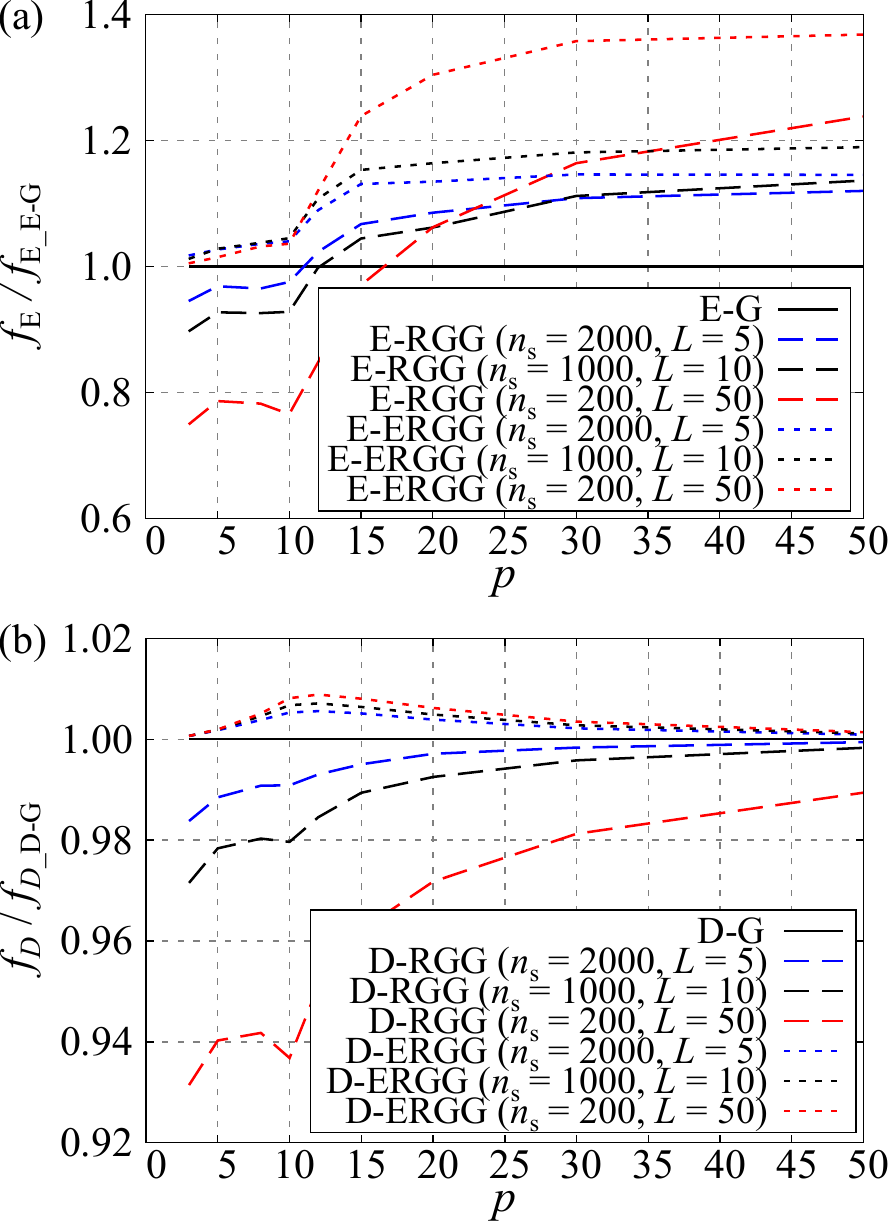}
\caption{Effects of the shrink ratio $n_{\rm s}/n$ on objective values of the randomized group-greedy methods. The number of evaluations for the RGG and ERGG methods are the same as that for the common greedy method ($L_{\rm RGG}n_{\rm s}=L_{\rm ERGG}n_{\rm s}=n$). (a) E-optimality-based methods; (b) D-optimality-based methods.}
\label{fig:eig-nr} 
\end{figure}

For oversampling conditions, the influence of $L$ on the objective value depends on the method. In the case of the E-GG method, there is no large effect of $L$ on the increase in objective value. On the other hand, the increase in objective values for the E-RGG and E-ERGG methods is larger than that for the E-GG method. The increase in the objective values for the E-RGG and E-ERGG methods is similar, that is, the difference of the objective value in the undersampling condition remains at larger $p$ when $L$ is the same. Even though the group size $L$ and the size of the sensor candidate matrix $n$ or $n_{\rm s}$ are different, the number of evaluations is the same for the same line colors in Fig.~\ref{fig:eig-L}. Although the objective value obtained by the E-RGG method is quite smaller than that obtained by the E-GG method under undersampling conditions, the objective values at $p>15$ obtained by the E-RGG and E-ERGG methods are larger than those obtained by the E-GG method when the number of evaluations is the same. In particular, the performance of the E-RGG and E-ERGG methods with $L=50$ is better than that of the E-GG method with $L=50$ at larger $p$, even though the number of evaluations for the E-GG method is ten times larger than that for the E-RGG and E-ERGG methods. Hence, it is considered that the group size is more important than the size of the sensor candidate matrix, unlike in undersampling conditions. This indicates that the combination of selected sensors is more important than the selected location itself in oversampling conditions. Consequently, the objective values obtained by the E-RGG and E-ERGG methods are considered to be improved by increasing $L$ instead of decreasing $n_{\rm s}$ in oversampling conditions. This is shown in Fig.~\ref{fig:eig-nr}. 

In the case of the D-optimality-based methods, the effect of $L$ on the objective value is smaller than that of the E-optimality-based methods. Even if $L$ for the D-RGG method increases until $L=500$, the objective value is smaller than that obtained by the D-GG method with $L=5$. Hence, the RGG method is not effective for the D-optimality-based greedy methods.

Since the randomized greedy method uses a shrunken sensor candidate matrix, the number of evaluations is a function of both the group $L$ size and the shrunken sensor candidate matrix $n_{\rm s}$. Fig.~\ref{fig:eig-nr} shows the influence of the size of the shrunken sensor candidate matrix on the objective value. It should be noted that the group size for the randomized methods was set so that the computational cost for the randomized methods would be the same as that for the common greedy method (i.e., $L_{\rm RGG}n_{\rm s}=L_{\rm ERGG}n_{\rm s}=n$). In the case of the E-RGG method, the objective value is lower than that obtained by the common greedy method around the undersampling condition and becomes small as the size of the shrunken sensor candidates $n_{\rm s}$ decreases. As the number of sensors increases, the objective value obtained by the E-RGG method becomes larger than that obtained by the common greedy method. In addition, the objective value obtained with smaller $n_{\rm s}$ (i.e., larger $L$) is larger, and the trend is different for smaller $p$. This trend indicates that the deeper search, i.e., larger $L$, is effective for oversampling conditions, even if $n_{\rm s}$ is small. For the E-ERGG method, the objective value in undersampling conditions is better than those obtained by the E-RGG method and the common greedy method because locations that are possibly valuable are secured by the elite strategy, and there is a small influence of $n_{\rm s}$. The increase in the objective values by increasing the number of sensors for E-RGG and E-ERGG methods under oversampling conditions is similar when $n_{\rm s}$ is the same. The difference between the objective values for these two methods is due to the influence of the difference in the sensors selected under undersampling conditions. The trend for the D-optimality-based methods is almost the same as that for E-optimality-based methods, but the D-RGG method cannot be superior to the common greedy method.

\section{Conclusions}
\label{sec:con}
The present study proposed randomized group-greedy methods. The E-optimality-based method and D-optimality-based method were implemented, and the performance of the proposed methods was compared with that for the common greedy method and the original group-greedy method. The performance evaluation was conducted by applying the methods to a randomly generated dataset in which the entries follow a normal distribution and have 10,000 potential sensor locations.

The results of the numerical experiment showed that the proposed method can obtain similar or better performance compared to the original group-greedy methods, while significantly reducing the computational cost. In particular, the proposed method is effective for the E-optimality-based method, in which the objective function does not have submodularity, and the performance is not high for the common greedy method.

When the number of evaluations for the objective function is the same as that for the common greedy method, the randomized group-greedy method is inferior to the common greedy method in performance under undersampling conditions but superior under oversampling conditions. By introducing the elite strategy, in exchange for a slight increase in computational cost, the common greedy method is surpassed in all conditions from the undersampling to the oversampling conditions. Furthermore, although the number for evaluations of the ERGG is lower than that for the original group-greedy method, it is possible to obtain the same level of performance under oversampling conditions as that for the original group-greedy method.

Since the computational cost of the proposed method is low, optimization with a larger group size is possible when the computational cost is the same as that for the original group-greedy method. In particular, the optimization results obtained by the proposed method are better than those obtained by the original group-greedy method when the number of sensors is large under oversampling conditions. The performance is further improved by including elite sensor candidates in the shrunken sensor candidate, and similar performance as the original group-greedy method can be obtained even under undersampling conditions.

For the D-optimality-based method, the objective function is a monotone submodular function. Because the degradation of the performance due to shrunken sensor candidates is significant, the proposed method was not effective. However, by including elite sensor candidates, it is possible to obtain similar or better performance compared to the original group-greedy method while reducing the computational cost, even for the D-optimality-based method. The proposed method is effective for problems involving a large number of sensor candidates, such as data-driven sensor selection problems.

\section*{Acknowledgments}
The present study was supported by JST CREST (JPMJCR1763), ACT-X (JPMJAX20AD), FOREST (JPMJFR202C), Japan and JSPS KAKENHI (JP21J20671), Japan. 

\bibliographystyle{IEEEtran}
\bibliography{main}

\begin{IEEEbiography}
[{\includegraphics[width=1in,height=1.25in,clip,keepaspectratio]{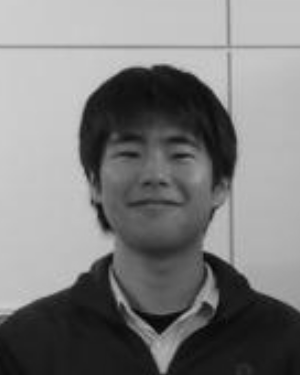}}]{Takayuki Nagata} received B.S. and M.S. degrees in mechanical and aerospace engineering from Tokai University, Hiratsuka, Japan, in 2015 and 2017, respectively. He received a Ph.D. degree in aerospace engineering from Tohoku University, Sendai, Japan, in 2020. From 2018 to 2020, he was a Research Fellow of the Japan Society for the Promotion of Science (JSPS) at Tohoku University, Japan. He is currently a postdoctoral researcher at Tohoku University, Sendai, Japan. 
\end{IEEEbiography}

\begin{IEEEbiography}
[{\includegraphics[width=1in,height=1.25in,clip,keepaspectratio]{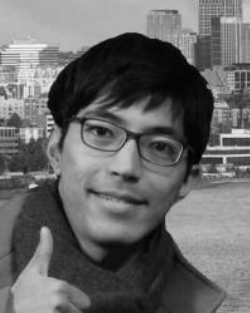}}]{Keigo Yamada} received a B.S. degree in physics from Tohoku University, Sendai, Japan, in 2019. He is currently a Ph.D. student at the Department of Aerospace Engineering at Tohoku University, Sendai, Japan. He is a Research Fellow of the Japan Society for the Promotion of Science (JSPS).
\end{IEEEbiography}

\begin{IEEEbiography}
[{\includegraphics[width=1in,height=1.25in,clip,keepaspectratio]{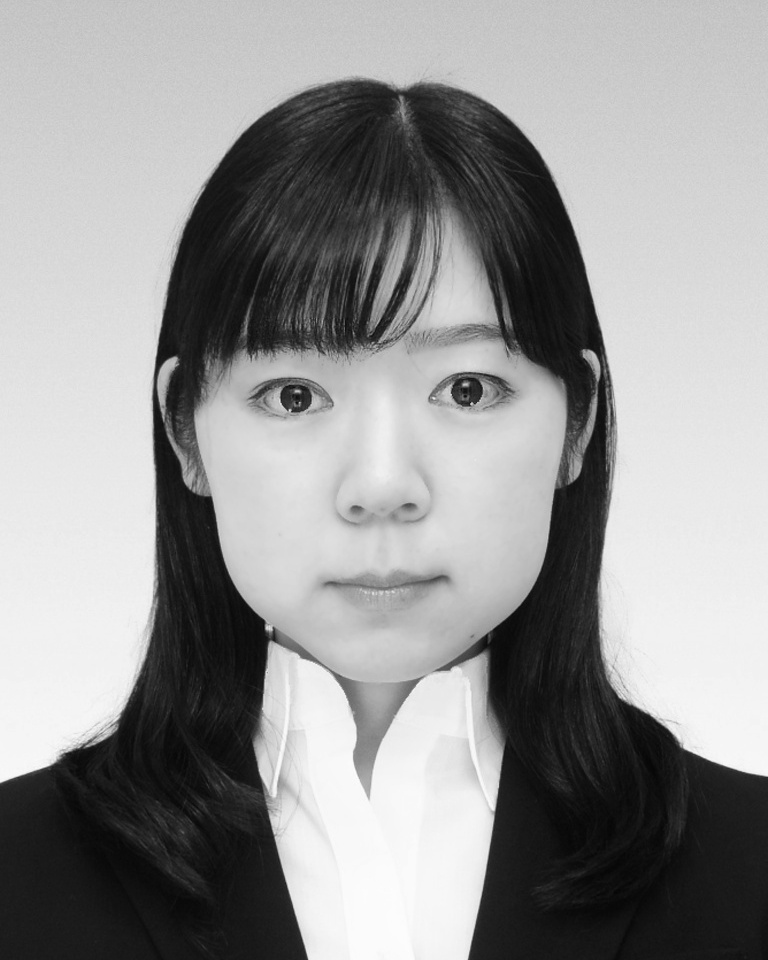}}]{Kumi Nakai} received a Ph.D. degree in mechanical systems engineering from Tokyo University of Agriculture and Technology, Tokyo, Japan, in 2020. From 2017 to 2020, she was a Research Fellow of the Japan Society for the Promotion of Science (JSPS) at Tokyo University of Agriculture and Technology, Tokyo, Japan. She is currently a postdoctoral researcher at Tohoku University, Sendai, Japan.
\end{IEEEbiography}

\begin{IEEEbiography}
[{\includegraphics[width=1in,height=1.25in,clip,keepaspectratio]{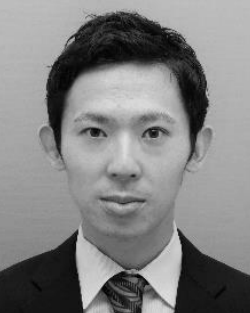}}]{Yuji Saito} received a B.S. degree in mechanical engineering, and a Ph.D. degree in mechanical space engineering from Hokkaido University, Japan, in 2018. He is currently an Assistant Professor at the Department of Aerospace Engineering at Tohoku University, Sendai, Japan.
\end{IEEEbiography}

\begin{IEEEbiography}
[{\includegraphics[width=1in,height=1.25in,clip,keepaspectratio]{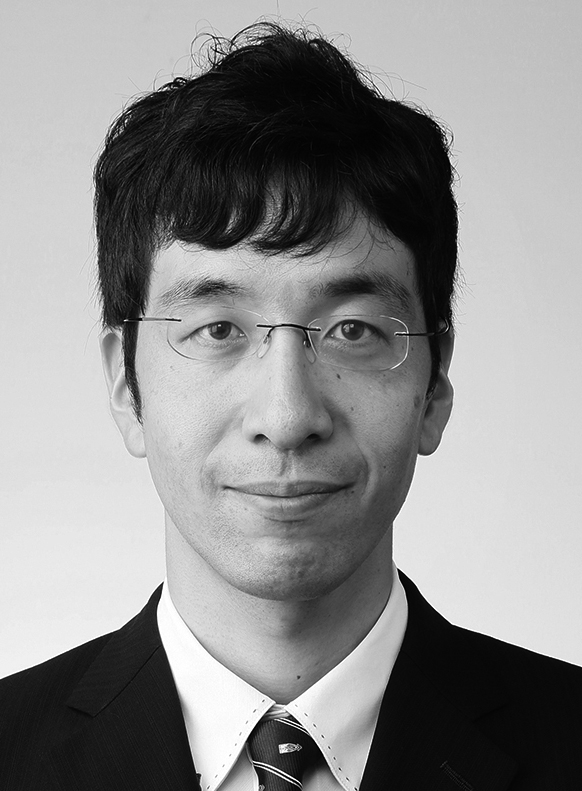}}]{Taku Nonomura} received a B.S. degree in mechanical and aerospace engineering from Nagoya University, Nagoya, Japan, in 2003, and a Ph.D. degree in aerospace engineering from the University of Tokyo, Japan in 2008. He is currently an Associate Professor at the Department of Aerospace Engineering at Tohoku University, Sendai, Japan.
\end{IEEEbiography}

\end{document}